\documentclass[12pt]{article}
\begin{document}
\newcommand {\sheptitle}
{Flavour leptogenesis with tribimaximal mixings and beyond}
\newcommand{\shepauthor}
{  H. Zeen Devi$^{\dag,*}$\footnote{Corresponding
    author.\\{\it{E-mail}}: zeenhijam@yahoo.co.in}, Amal Kr
  Sarma$^{\ddag}$  and  N. Nimai Singh$^{\dag}$ }
\newcommand{\shepaddress}
{$^{\dag}$ Department of Physics,Gauhati University,Guwahati-781014, India \\
$^{\ddag}$ Department of Physics, D.R College, Golaghat, Assam, India}
\newcommand{\shepabstract}
 {We compute and compare the baryon asymmetry of the universe in thermal 
 leptogenesis scenario with and without flavour effects for different neutrino 
mass models namely degenerate, inverted hierarchical and normal hierarchical
 models, with tribimaximal mixings and beyond. Considering three possible 
 diagonal forms of Dirac neutrino mass matrices $m_{LR}$, the right-handed
 Majorana mass matrices $M_{RR}$ are constructed from the light neutrino
 mass matrices $m_{LL}$ through the inverse seesaw formula. The normal
 hierarchical model is found to give the best  predictions of the baryon 
 asymmetry for both cases. This 
  analysis serves as an additional information in the discrimination of the 
  presently available neutrino mass models. Moreover, the flavour effects is found to give enhancement of the baryon asymmetry in thermal leptogenesis.} 

\begin{titlepage}
\begin{flushright}
\end{flushright}
\begin{center}
{\large{\bf\sheptitle}}
\bigskip \\
\shepauthor
\\
\mbox{}
\\
{\it \shepaddress}
\\
\vspace{.5in}
{\bf Abstract} \bigskip \end{center}\setcounter{page}{0}
\shepabstract
\end{titlepage}
\section{Introduction}
The existence of heavy right-handed Majorana  neutrinos in some of the
 left-right symmetric GUT models,  not only gives small but
 non-vanishing
 neutrino masses through the celebrated seesaw mechanism\cite{ref SSW1}, 
 it also  plays an important role in explaining the baryon asymmetry
 of the universe \cite{ref DNN2,ref Dunkley:2008ie}. Such an asymmetry can be dynamically generated if the particle interaction rate and the  expansion rate of the universe satisfy Sakharov's three famous conditions \cite{ref SKH3}. Majorana right-handed neutrinos satisfy the second condition i.e., C and CP violation as they can have an asymmetric decay to leptons and Higgs particles, and the process occurs at different rates for particles and antiparticles. The lepton asymmetry is then partially converted to baryon asymmetry by electroweak sphaleron process \cite{ref KRS4,ref YF5,ref MAL92,ref KTE,ref US98,ref DF02}.

\vspace{0.5cm}
\indent
In order to calculate the baryon asymmetry from a given neutrino mass
model, one usually  starts with the light neutrino mass matrices $m_{LL}$
 and then relates it with  the heavy Majorana neutrino mass matrices $M_{RR}$ 
 and the Dirac neutrino mass  matrix $m_{LR}$ through the inverse seesaw 
mechanism in an elegant way. We consider  the Dirac neutrino mass 
 matrix $m_{LR}$ as either the  charged lepton mass matrix, down-quark
 mass
 matrix or up-quark mass matrix for phenomenological analysis.
 The complex CP violating phases necessary for lepton asymmetry are usually derived from the MNS leptonic
mixing matrix. In the present work we are interested to  consider the
complex
 Majorana phases which are derived from the right-handed Majorana mass
 matrix
 $M_{RR}$, in the estimation of baryon asymmetry of the universe.
 We consider the left-handed light Majorana neutrino mass
 matrices
 $ m_{LL}$ which obey the $\mu-\tau$ symmetry, where tribimaximal
 mixings and below are realised \cite{ref NZM20,ref deg}, for all possible patterns of neutrino masses, 
viz, degenerate,
 inverted hierarchical and normal hierarchical mass patterns. 
 We first parametrise the light left-handed Majorana neutrino mass
 matrices which are subjected to correct predictions of neutrino mass
 parameters and mixing angles.  The calculation of baryon asymmetry
 may serve as an additional information to further discriminate the
 correct
 pattern of neutrino mass models and also shed light on the structure
 of Dirac neutrino mass matrix.

\vspace{0.5cm}   
 \indent
In section 2 we briefly mention the formalism for estimating the
 lepton 
asymmetry in flavoured thermal leptogenesis through the
``out-of-equilibrium'' 
decay of the heavy right-handed Majorana neutrinos and also discuss briefly on $\mu-\tau$ symmetry with Tribimaximal mixings(TBM) as a special case. Section 3 is
devoted
 to  the numerical calculation and results. Finally in section 4 we
 conclude with  a summary and discussions. Important expressions
 related
 to $m_{LL}$ which obey $\mu-\tau$ symmetry for three neutrino mass
 models,
 are relegated to Appendix A.

\section{Flavoured Thermal leptogenesis}
The canonical seesaw formula \cite{ref SSW1} relates the left-handed Majorana  
neutrino mass matrix $m_{LL}$ 
and heavy right handed Majorana mass matrix $M_{RR}$ in a  simple way
\begin{equation}\label{ch301}
m_{LL} = -m_{LR}M_{RR}^{-1}m_{LR}^{T}
\end{equation}
where  $m_{LR}$ is the Dirac neutrino mass matrix. 
For our calculation of lepton asymmetry, we consider the model\cite{ref KRS4,ref YF5,ref MAL92}
 where the asymmetric decay of the lightest of the heavy right-handed Majorana neutrinos,
 is assumed.  The physical  Majorana neutrino $N_{R}$  decays into two modes:
\begin{center}
$N_{R}\rightarrow l_{L}+\phi^{\dagger}$\\

      $\rightarrow \overline{l}_{L}+\phi$
\end{center}
where  $l_{L}$ is the lepton and $\bar{l}_{L}$ is the antilepton and  the
 branching ratio for these two decay modes is  likely to be different. 
The CP-asymmetry which is caused by the intereference of tree level with one-loop corrections
 for the decays of lightest of heavy right-handed Majorana neutrino $N_{1}$, is defined by \cite{ref YF5,ref KTE}
\begin{center}
$\epsilon=\frac{\Gamma -\overline{\Gamma}}{\Gamma+\overline{\Gamma}}$
\end{center}
where  $\Gamma=\Gamma(N_{1}\rightarrow l_{L}\phi^{\dagger})$ 
and $\overline{\Gamma}=\Gamma(N_{1}\rightarrow \overline{l_{L}}\phi)$
are the decay rates.

\vspace{0.5cm}
\indent
In this section we study the flavour effects in leptogenesis \cite{ref DNN2} in the context of our neutrino mass models in \cite{ref NZM20,ref deg} (see Appendix A for details). Earlier leptogenesis calculations were done by studying Boltzman Equations (BE) for the B-L asymmtery. But
 later \cite{ref BCSN22} studied flavour $B-L_{\alpha}$ asymmetries where the
 results were significantly different from the "single flavour approximation". 
Subsequently many authors \cite{ref AAL23,ref NT24,ref OV25} have included flavour effects to enhance the baryon asymmetry in particular models. In thermal leptogenesis the importance of flavour effects comes from the wash-out effects, where scattering produces $N_1$ population of neutrinos at temperature $T\simeq M_1$. When T drops below $M_1$, this $N_1$ population decays to leptons and if these decays are CP violating, it can produce asymmetries in all the lepton flavours. If the interactions are "out-of-equillibrium", then the above asymmetries would survive.\\ \\
\indent
 In thermal leptogenesis \cite{ref DNN2} the Yukawa coupling constant  related 
to the production of $N_1$ also controls the decay of $N_1$. Initially it
seems that both the CP asymmetries will be washed out leaving no lepton
 asymmetry. 
However, a net asymmetry survives after the potential
 cancellation of CP asymmetry between processes with $N_1$ and
 $l_\alpha$($\alpha=e,\mu, \tau$)
 in the final state such as $X\rightarrow Nl_\alpha$ scattering and
 N in the initial state and $l_\alpha$ in the final state, such as 
$N \rightarrow \phi l_\alpha$. Only processes with $l_\alpha$ in the
 final state can produce the asymmetry. There is no cancallation
 between
 asymmetries produced in the decays and inverse decays. Any initial
 asymmetry produced with the N population is depleted by scattering, decays and inverse decays. This depleton is called washout. The
initial
 state of washout contains a lepton, so it is important to know which
 leptons are distinguishable. It is always assumed that interactions
 whose
 timescale is very different from the leptogenesis scale are dropped 
out from the Boltzman Equations(BE) \cite{ref Pilaftsis,ref Abada1,ref Nardi,ref Abada2}.

In the interaction Lagrangian the
different
 flavours are disinguished by their Yukawa couplings $h_\alpha$. Thus
 if 
the $h_\alpha $ mediated interactions are fast compared to the
leptogenesis
 scale and the universe expansion rate, these distinguishable 
$h_\alpha$ will have induced differences in the thermal masses of
 different leptons as each of $h_{e,\mu ,\tau} $ has different
 strengths.
 Thus when the charge lepton Yukawa interactions are fast then flavour
 basis is the correct basis for the BE, otherwise leptogenesis has no
 knowledge of the lepton flavour for 'slow interactions'.\\ \\

 In the flavour basis the equation for the lepton asymmetry in $N_{1} $ decay 
becomes$(\alpha=e,\mu,\tau)$,
\begin{equation}\label{ch302}
\epsilon_{\alpha\alpha} = \frac{1}{8\pi}\frac{1}{(h^{\dag}h)_{11}}\left[ \sum_{j=2,3} Im\left[ h^{*}_{\alpha 1}(h^{\dag}h)_{1 j}h_{\alpha j}\right]g(x_j)
+\sum_j Im\left[ h^{*}_{\alpha 1}(h^{\dag}h)_{j 1}h_{\alpha j}\right] \frac{1}{(1-x_j)} \right]
\end{equation}
\begin{equation}\label{ch303}
x_j\equiv\frac{M^{2}_j}{M^{2}_i} ; \ \ \
g(x_j)\sim\frac{3}{2}\frac{1}{\sqrt(x_j)}. 
\end{equation}
The efficiency factor for ``out-of-equilibrium'' situation
 i.e., $\Gamma_{ID}<H=1.66\sqrt{g_*}\frac{T^2}{m_{Pl}}$  is given by\cite{ref DNN2}
\begin{equation}\label{ch304}
\eta_{\alpha}\equiv\frac{m_{*}}{\tilde{m}_{\alpha\alpha}}
\end{equation}
where $m_{*}=8\pi\frac{v^2}{M_{1}^{2}}H\sim 1.1\times 10^{-3}eV $ \cite{ref DNN2} and
\begin{equation}\label{ch305}
\tilde{m}=\frac{h^{\dag}_{\alpha 1}h_{\alpha 1}v^{2}}{M_1}.
\end{equation}
For SM we have
\begin{equation}\label{ch306}
v= 174GeV, \ \ \ g_*=106.75 .
\end{equation}
The second term in Eq.[\ref{ch302}] violates the single lepton flavours
but
 conserves the total lepton number. It vanishes when summed over
 flavours. Thus,
\begin{equation}\label{ch307}
\epsilon\equiv\sum_{\alpha}\epsilon_{\alpha\alpha}=\frac{1}{8\pi}\frac{1}{(h^{\dag}h)_{11}}\sum_{j} Im\left[ {(h^{\dag}h)}^{2}_{1j} \right]g(x_j).
\end{equation}
Thus in the strong washout case for all flavours, we obtain the baryon
asymmetry i.e; baryon-to-entropy \cite{ref DNN2} ratio as, 
\begin{equation}\label{ch308}
Y_{3B}\sim10^{-3}\sum_{\alpha}\eta_{\alpha}\epsilon_{\alpha\alpha}\sim10^{-3}m_{*}\sum_{\alpha}\frac{\epsilon_{\alpha\alpha}}{\tilde{m_{\alpha\alpha}}},  
\end{equation} 
For single flavoured case, one can consider the direction in flavour 
space into which $N_{1}$ decays. In single flavour case the baryon asmmetry is given by
\begin{equation}\label{ch309}
Y_{1B}\sim 10^{-3}m_{*}\frac{\epsilon}{\tilde{m}},
\end{equation}
where
\begin{equation}\label{ch310}
\epsilon=\sum_{\alpha}\epsilon_{\alpha\alpha}, \ \ \ \tilde{m}=\sum_{\alpha}\tilde{m}_{\alpha\alpha},
\end{equation} 
The entropy density $s$ of the universe can be related to the photon number density $\eta_{\gamma}$ as $s=7.04\eta_{\gamma}$ . So baryon-to-entropy ratio is estimated to be around $\sim 8.74\times10^{-11}$ \cite{ref Dunkley:2008ie}. 
\subsection{Neutrino mass models with $\mu-\tau$ symmetry: Tribimaximal mixings}
The recent global 3$\nu$ oscillation analysis \cite{ref Gracia} indicates towards a specific form of leptonic mixing - Tribimaximal mixing and a slight deviation from tribimaximal mixing pattern which is a special case of $\mu-\tau$ symmetry. The $\mu-\tau$ reflection symmetry in the neutrino mass matrix, implies an invariance under the simultaneous
permutation of the second and third rows as well as the second and
third columns in neutrino mass matrices \cite{ref HS6,ref Ahn7,ref koide8,ref koide9,ref Moha10,ref HPS11,ref Ma12},
\begin{equation}
 m_{LL}=
 \left(\begin{array}{ccc}
 X  &  Y  &   Y\\
  Y   &  Z  &   W \\
 Y  &  W  &  Z
 \end{array}\right).
 \end{equation}
 This has the  permutation symmetry, $Pm_{LL}P=m_{LL}$, where
\begin{equation}
 P=
 \left(\begin{array}{ccc}
 1 &  0  &   0 \\
  0  &  0  &   1 \\
 0  &  1  &  0
 \end{array}\right).
 \end{equation}
Neutrino mass matrix in eq.(11) predicts the  maximal atmospheric
mixing angle,
$\theta_{23}=\pi/4$ and  $\theta_{13}=0$. However the prediction on
solar mixing 
angle $\theta_{12}$ is arbitrary, and  it can be fixed by  the input values of the
parameters present in the mass matrix. Thus
\begin{equation}
\tan2\theta_{12}=|\frac{2\sqrt{2}Y}{(X-Z-W)}|
\end{equation}
which  depends on four input parameters
$X,Y,Z$ and $W$. This makes us difficult to choose the values of these
free parameters for a  solution consistent with neutrino oscillation
data. This point is addressed in \cite{ref NZM20,ref deg} where the solar
angle is made dependent  only on  the ratio of two parameters, $\eta/\epsilon$. Such parametrization of the mass matrix enables us to analyse the neutrino
mass  matrix in a  systematic and economical way \cite{ref NMA13}. The actual
values of these two new parameters will be fixed by the data on neutrino mass
squared differences. 
 
The MNS leptonic mixing matrix $U_{MNS}$ which diagonalises  $m_{LL}$
is defined by
$m_{LL}=U_{MNS}DU^{\dagger}_{MNS}$  where $D=diag.(m_1, m_2, m_3)$, and 
 \begin{equation}
 U_{MNS}=
 \left(\begin{array}{ccc}
 U_{e1} &  U_{e2}  &   U_{e3} \\
  U_{\mu 1}  &  U_{\mu 2}  &   U_{\mu 3} \\
 U_{\tau 1}  &  U_{\tau 2}  &  U_{\tau 3}
 \end{array}\right).
 \end{equation}
 From the  consideration of $\mu$-$\tau$ reflection symmetry,
 $U_{MNS}$ has  the following general  properties \cite{ref HS6},
$|U_{\mu i}|=|U_{\tau i}|$, $|U_{\mu i}|^2=(1-|U_{ei}|^2)/2$
where  $i=1,2,3$. For $i=3$, 
$|U_{\mu 3}|^2=(1-|U_{e3}|^2)/2$.
For $|U_{e3}|=0$, we have $|U_{\mu 3}|=|U_{\tau 3}|=1/\sqrt{2}$. 
The MNS mixing matrix is generally  parametrised by three rotations
($\theta_{23}=\pi/4$, $\theta_{13}=0$): 
\begin{equation}
 U_{MNS}=O_{23}O_{13}O_{12}=O_{23}O_{12}=
 \left(\begin{array}{ccc}
 c_{12} &  -s_{12}  &   0 \\
  s_{12}/\sqrt{2}  &  c_{12}/\sqrt{2}  &  - 1/\sqrt{2} \\
 s_{12}/\sqrt{2}  &  c_{12}/\sqrt{2}  &  1/\sqrt{2}
 \end{array}\right)
 \end{equation}
where $c_{12}=\cos\theta_{12}$,
$s_{12}=\sin\theta_{12}$. Tri-bimaximal mixing (TBM) is a special case with
$c_{12}=\sqrt{2/3}$ and $s_{12}=\sqrt{1/3}$ \cite{ref HPS11,ref Ma12},
\begin{equation}
 U_{TBM}=
 \left(\begin{array}{ccc}
 \sqrt{2/3} &  -1/\sqrt{3}  &   0 \\
  1/\sqrt{6}  &  1/\sqrt{3}  &  - 1/\sqrt{2} \\
 1/\sqrt{6}  &  1/\sqrt{3}  &  1/\sqrt{2}
 \end{array}\right)
 \end{equation}
where
\begin{equation}
 O_{23}=
 \left(\begin{array}{ccc}
 0 &  0  &   0 \\
 0  &  1/\sqrt{2}  &  - 1/\sqrt{2} \\
 0  &  1/\sqrt{2}  &  1/\sqrt{2}
 \end{array}\right),
 \end{equation}
and
\begin{equation}
 O_{12}=
 \left(\begin{array}{ccc}
 \sqrt{2/3} &  -1/\sqrt{3}  &   0 \\
  1/\sqrt{3}  &  \sqrt{2/3}  &  0 \\
 0  &  0  &  1
 \end{array}\right).
 \end{equation}
For completeness we also give the three neutrino  mass eigenvalues\cite{ref koide9}
corresponding to  the
neutrino mass  matrix in eq.(11),
\begin{equation}
-m_1=\frac{1}{2}[Z+W+X-\sqrt{8Y^2+(Z+W-X)^2}],
\end{equation}
\begin{equation}
m_2=\frac{1}{2}[Z+W+X+\sqrt{8Y^2+(Z+W-X)^2}],
\end{equation}
\begin{equation}
m_3=(Z-W).
\end{equation}
The solar mixing  angle is given by 
$\cos\theta_{12}=\sqrt{\frac{m_2+X}{m_1+m_2}}$, $\sin\theta_{12}=\sqrt{\frac{m_1-X}{m_1+m_2}}$.
If $X=0$, then we have a simple relation, $\tan^2\theta_{12}=m_1/m_2$. 
For Tribimaximal mixing we get $\tan^2\theta_{12}=0.5$ , $\tan^2\theta_{23}=1$ and  $\tan^2\theta_{13}=0$ for particular 'flavor twister' term as mentioned in details in our earlier works \cite{ref NZM20, ref NMA13}. 
\section{Numerical estimation of baryon asymmetry}
For numerical calculation we first choose the light
left-handed Majorana
 neurino mass matrix $m_{LL}$ proposed in
Appendix A \cite{ref NZM20,ref deg}. These mass matrices obey the $\mu-\tau$ symmetry which guarantees the tribimaximal mixings.

 The Dirac neutrino mass matrix $m_{LR}$ appeared in seesaw formula can have any arbitrary   structure which either is   diagonal or non-diagonal. In the seesaw mechanism, for a specific structure of $m^I_{LL}$, we can have three possible combinations of $m_{LR}$ and $M_{RR}$:
(a) both $m_{LR}$ and $M_{RR}$ are non-diagonal, (b) $m_{LR}$ diagonal and $M_{RR}$ non-diagonal,
(c) $m_{LR}$ non-diagonal and $M_{RR}$ diagonal.
These three combinations  can be realised in different physical situations. For example, when one calculate lepton asymmetry, one needs to consider the diagonal basis of heavy right-handed neutrinos, and combination (c) becomes  relevant. 
In the present calculations, the  diagonal form of $m_{LR}$ is chosen for different  neutrino  mass matrices.To see this let us consider the seesaw relation 
$m^I_{LL}= - m_{LR} M^{-1}_{R}m_{LR}^T$, where both $m_{LR}$ and $M_R$ are non-diagonal. 
Using some left and right handed rotations, the Dirac neutrino mass matrix can be diagonalised \cite{ref AFS14} as 
\begin{equation}\label{ch311}
m^{diag}_{LR} = U_L m_{LR} U^{\dag}_R.
\end{equation}
In terms of diagonal basis of  $m_{LR}$, the seesaw relation reduces to  
\begin{equation}\label{ch312}
m^I_{LL} = -U_L^{\dag} m_{LR}^{diag} M_{RR}^{-1} m_{LR}^{diag} U_L^*,
\end{equation}
where, $M^{-1}_{RR} = U_R M^{-1}_R U^T_R $.
It is assumed that eigenvalues of $m_{LR}^{diag}$ are hierarchical (similar to quarks or charged leptons). In absence of Dirac left handed rotations \cite{ref AFS14}, we can set $U_L \sim 1$. For slight deviation from unity we can assume $U_L \simeq  U_{CKM}$, where $U_{CKM}$ is the  quark mixing matrix. Again this can be set to unity  as quarks mixings are very small. This type of approximations do not produce significant change in numerical calculations.  For $U_L \sim 1$ the eq. (\ref{ch312})reduces to $m^I_{LL} = -m_{LR}^{diag} M_{RR}^{-1} m_{LR}^{diag}$ where  $M_{RR}$ is in the diagonal basis of $m_{LR}$. We follow this representation in the present calculation.\\
In some Grand Unified Theory such as $SO(10)$ GUT , the possible structure of $m_{LR}$ \cite{ref FK16} can be $ m_{LR} = diag(\lambda^m,\lambda^n,1)v$ , where $v$ is the overall scale factor representing electroweak vacuum expectation values. In the  present calculation we take  $\lambda =0.3$ and $v =174 GeV$. We consider three choices  of $(m,n)$ pair: case(i)  $(m,n)\equiv (6,2)$ for charged lepton, (ii)  $(m,n)\equiv (8,4)$ for  up-quark mass matrices and (iii)$(m,n)\equiv (4,2)$ for down-quark mass matrices representing  the  Dirac neutrino mass matrix.\\ 
 
For our calculation we choose a basis $U_{R}$ where 
$M_{RR}^{diag} = U_{R}^{T} M_{RR} U_{R}$
=diag($M_{1},M_{2},M_{3}$)
 with real and positive eigenvalues \cite{ref AFS14,ref AZN15}.
In this prime basis the Dirac neutrino Yukawa coupling becomes  
$h =\frac{m'_{LR}}{v}$ \cite{ref AFS14}which enters in the expression of CP-asymmetry $\epsilon_{\alpha\alpha}$ in Eq.[\ref{ch302}]. The term $h^{\dag}h$ is in the basis where the $M_{R}$ is diagonal with real and positive eigenvalues. Using the relation $h=m_{LR}/v$ and  Eq.[\ref{ch311}] we get
\begin{equation}\label{ch13}
h^{\dag}h=\frac{1}{v^2} U^{\dag}{R}(m_{LR}^{dia})^{2}U_{R},
\end{equation}
where v is the electroweak vacuum expectation value(174 GeV).\\
By this we allow the non-zero elements $M_{i}$ of the diagonalised RH neutrino mass matrix $M^{dia}_{R}$ to be complex. The unitary matrix $U_{R}$ is defined in such a way that it relates the basis where $m_{LR}$ is diagonal to the basis where $M_{R}$ is diagonal with real and positive non-zero elements.i.e, the phases of $M_{i}$ should be included in the defination of $U_{R}$.\\
So, we transform 
$m_{LR}=diag(\lambda^m, \lambda^n, 1)v$ to the $U_R$ basis by 
$m_{LR}\rightarrow m'_{LR} U_R$.

The Yukawa coupling matrix  $h = \frac{m'_{LR}}{v}$ so constructed, also becomes complex, and hence the term $Im( h^{\dag}h)_{1j}$ appearing in lepton asymmetry $\epsilon_{\alpha\alpha}$ gives a non-zero contribution. In our numerical estimation of lepton asymmetry, we choose some arbitrary values of $\alpha$ and $\beta$ other than $\pi/2$ and $0$. For example, light neutrino masses $(m_1, -m_2, m_3)$ lead to $M_{RR}^{diag}= diag(M_1, -M_2, M_3)$,and we thus fix the Majorana phase $Q = diag ( 1,e^{(i \alpha)}, e^{(i \beta)}) 
= diag(1,e^{i( \pi/2 +\pi/4)},e^{i\pi/4})$ for 
$\alpha =(\pi/4 +\pi/2)$ and $\beta =\pi/4$. 
The extra phase $\pi/2$ in $\alpha$ absorbs the negative sign before heavy Majorana mass $M_2$. In our search programme such choice of the phases leads to highest numerical estimations of lepton CP asymmetry. The corresponding light left-handed neutrino mass matrix obeying $\mu-\tau$ symmetry, is collected from Appendix A as mentioned.
\begin{table}[tbp]
\begin{tabular}{cccccc}\hline
Type&$\Delta m^{2}_{21}[10^{-5}eV^{2}]$&$\Delta m^{2}_{23}[10^{-3}eV^{2}]$&$\tan^{2}\theta_{12}$&$\sin^{2}2\theta_{23}$&$\sin\theta_{13}$\\
\hline
 Deg.(IA) &7.8&2.6&0.5&1.0&0.0\\
 Deg.(IB) &7.9&2.5&0.5&1.0&0.0\\
 Deg.(IC) &7.9&2.5&0.5&1.0&0.0\\
\hline
 IH.(IIA) &7.3&2.5&0.5&1.0&0.0\\
 IH.(IIB) &8.5&2.3&0.5&1.0&0.0\\
\hline
NH.(IIIA) &7.1&2.1&0.5&1.0&0.0\\
NH.(IIIB) &7.5&2.4&0.5&1.0&0.0\\
\hline
\end{tabular}
\hfil
\caption{\footnotesize  Predicted values of the  solar and atmospheric neutrino
mass-squared differences  for $\tan^{2}\theta_{12}$=0.50, using  $m_{LL}$  given in the
Appendix A}

\end{table}
\begin{table}[tbp]
\begin{tabular}{cccccc}\hline
Type&$\Delta m^{2}_{21}[10^{-5}eV^{2}]$&$\Delta m^{2}_{23}[10^{-3}eV^{2}]$&$\tan^{2}\theta_{12}$&$\sin^{2}2\theta_{23}$&$\sin\theta_{13}$\\
\hline
 Deg.(IA) &7.6&2.6&0.45&1.0&0.0\\
 Deg.(IB) &7.9&2.8&0.45&1.0&0.0\\
 Deg.(IC) &7.9&2.5&0.45&1.0&0.0\\
\hline
 Inh.(IIA) &7.6&2.5&0.45&1.0&0.0\\
 Inh.(IIB) &8.4&2.0&0.45&1.0&0.0\\
\hline
Nh.(IIIA) &7.7&2.6&0.45&1.0&0.0\\
Nh.(IIIB) &8.0&2.6&0.45&1.0&0.0\\
\hline
\end{tabular}
\hfil
\caption{\footnotesize  Predicted values of the  solar and atmospheric neutrino
mass-squared differences  for $\tan^{2}\theta_{12}$=0.45, using  $m_{LL}$  given in the
Appendix A.   }
\end{table}
\begin{table}[tbp]
 \begin{tabular}{l l l l l} \hline
Type & (m,n) & $ M_{1}$GeV & $ M_{2}$GeV & $ M_{3}$GeV \\ \hline
IA & (4,2) & 1.46 $ \times$ $10^{10}$ & -6.20 $\times$ $10^{11}$ & 2.59 $\times$ $10^{13}$\\
IA & (6,2) & 1.22 $ \times$ $10^{8}$ & -6.01 $\times$ $10^{11}$ & 2.59 $\times$ $10^{13}$\\
IA & (8,4) & 9.86 $ \times$ $10^{5}$ & -5.03 $\times$ $10^{9}$ & 2.51 $\times$ $10^{13}$\\ 
IB & (4,2) & 5.01 $\times$ $10^{9}$ & 6.16 $\times$ $10^{11}$ & 7.60 $\times$ $10^{13}$\\
IB & (6,2) & 4.05 $\times$ $10^{7}$ & 6.16 $\times$ $10^{11}$ & 7.60 $\times$ $10^{13}$\\
IB & (8,4) & 3.28 $\times$ $10^{5}$ & 4.99 $\times$ $10^{9}$ & 7.60 $\times$ $10^{13}$\\ 
IC & (4,2) & 5.01 $\times$ $10^{9}$ & -6.69 $\times$ $10^{12}$ & 6.99 $\times$ $10^{12}$\\
IC & (6,2) & 4.05 $\times$ $10^{7}$ & -6.69 $\times$ $10^{12}$ & 6.99 $\times$ $10^{12}$\\
IC & (8,4) & 3.28 $\times$ $10^{5}$ & -4.83 $\times$ $10^{11}$ & 7.84 $\times$ $10^{11}$\\ \hline 
IIA & (4,2)&4.01$\times$ $10^{10}$& 9.73$\times$$10^{12}$&6.25$\times$$10^{16}$\\
IIA & (6,2)&3.29$\times$ $10^{8}$& 9.73$\times$$10^{12}$&6.25$\times$$10^{16}$\\
IIA & (8,4)&2.63$\times$ $10^{6}$& 7.94$\times$$10^{10}$&6.21$\times$$10^{16}$\\
IIB & (4,2)&-1.19$\times$ $10^{11}$& 2.71$\times$$10^{12}$&5.59$\times$$10^{14}$\\
IIB & (6,2)&-9.97$\times$ $10^{8}$& 2.63$\times$$10^{12}$&5.59$\times$$10^{14}$\\
IIB & (8,4)&-8.10$\times$ $10^{6}$& 2.14$\times$$10^{10}$&5.57$\times$$10^{14}$\\
 \hline 
IIIA & (4,2) &3.59$\times$ $10^{12}$&-5.48$\times$$10^{12}$&2.89$\times$ $10^{14}$\\
IIIA & (6,2)& 3.93$\times$ $10^{11}$&-4.09$\times$ $10^{11}$&2.87$\times$ $10^{14}$\\
IIIA & (8,4) &3.19$\times$ $10^{9}$&-3.22$\times$ $10^{9}$&2.85$\times$ $10^{14}$\\
IIIB & (4,2) &3.57$\times$ $10^{12}$&-5.29$\times$$10^{12}$&3.01$\times$ $10^{14}$\\
IIIB & (6,2)& 3.85$\times$ $10^{11}$&-3.99$\times$ $10^{11}$&2.99$\times$ $10^{14}$\\
IIIB & (8,4) &3.13$\times$ $10^{9}$&-3.25$\times$ $10^{9}$&2.97$\times$ $10^{14}$\\
\hline
\end{tabular}
\hfil
\caption{\footnotesize  Heavy right-handed  Majorana neutrino  masses  $M_{j}$
 for  degenerate models (IA,IB,IC), inverted models (IIA,IIB)
and normal hierarchical models (IIIA, IIIB),  with
$\tan^2\theta_{12}$=0.5, using light neutrino mass matrices $m_{LL}$ given in
Appendix A. The entry $(m,n)$ indicates the type of Dirac neutrino
mass matrix $m_{LR}=(\lambda^m,\lambda^n,1)v$, as down quark mass matrix $(4,2)$, charged lepton mass
matrix (6,2) and up quark mass matrix $(8,4)$, as explained in the text.  }
\end{table}
\begin{table}[tbp]
 \begin{tabular}{l l l l l} \hline
Type & (m,n) & $ M_{1}$ & $ M_{2}$ & $ M_{3}$ \\ \hline
IA & (4,2) & 5.43 $ \times$ $10^{10}$ & -3.34 $\times$ $10^{12}$ & 8.42 $\times$ $10^{13}$\\
IA & (6,2) & 4.51 $ \times$ $10^{8}$ & -3.26 $\times$ $10^{12}$ & 8.42 $\times$ $10^{13}$\\
IA & (8,4) & 3.65 $ \times$ $10^{6}$ & -2.77 $\times$ $10^{10}$ & 8.03 $\times$ $10^{13}$\\ 
IB & (4,2) & 5.01 $\times$ $10^{9}$ & 6.16 $\times$ $10^{11}$ & 7.60 $\times$ $10^{13}$\\
IB & (6,2) & 4.05 $\times$ $10^{7}$ & 6.16 $\times$ $10^{11}$ & 7.60 $\times$ $10^{13}$\\
IB & (8,4) & 3.28 $\times$ $10^{5}$ & 4.99 $\times$ $10^{9}$ & 7.60 $\times$ $10^{13}$\\ 
IC & (4,2) & 5.01 $\times$ $10^{9}$ & -6.69 $\times$ $10^{12}$ & 6.99 $\times$ $10^{12}$\\
IC & (6,2) & 4.05 $\times$ $10^{7}$ & -6.69 $\times$ $10^{12}$ & 6.99 $\times$ $10^{12}$\\
IC & (8,4) & 3.28 $\times$ $10^{5}$ & -4.81 $\times$ $10^{11}$ & 7.86 $\times$ $10^{11}$\\ \hline
IIA & (4,2)&4.02$\times$ $10^{10}$& 9.73$\times$$10^{12}$&6.59$\times$$10^{16}$\\
IIA & (6,2)&3.25$\times$ $10^{8}$& 9.73$\times$$10^{12}$&6.59$\times$$10^{16}$\\
IIA & (8,4)&2.63$\times$ $10^{6}$& 7.94$\times$$10^{10}$&6.54$\times$$10^{16}$\\
IIB & (4,2)&-9.76$\times$ $10^{10}$& 2.89$\times$$10^{12}$&6.23$\times$$10^{14}$\\
IIB & (6,2)&-8.10$\times$ $10^{8}$& 2.83$\times$$10^{12}$&6.23$\times$$10^{14}$\\
IIB & (8,4)&-6.56$\times$ $10^{6}$& 2.29$\times$$10^{10}$&6.21$\times$$10^{14}$\\
 \hline  
IIIA & (4,2) &1.74$\times$ $10^{12}$&-2.28$\times$$10^{13}$&2.96$\times$ $10^{14}$\\
IIIA & (6,2)& 1.83$\times$ $10^{10}$&-1.82$\times$ $10^{13}$&1.04$\times$ $10^{14}$\\
IIIA & (8,4) &1.48$\times$ $10^{8}$&-1.79$\times$ $10^{11}$&8.56$\times$ $10^{13}$\\
IIIB & (4,2) &3.73$\times$ $10^{12}$&-5.68$\times$$10^{12}$&2.96$\times$ $10^{14}$\\
IIIB & (6,2)& 4.08$\times$ $10^{11}$&-4.24$\times$ $10^{11}$&2.93$\times$ $10^{14}$\\
IIIB & (8,4) &3.31$\times$ $10^{9}$&-3.45$\times$ $10^{9}$&2.91$\times$ $10^{14}$\\
\hline
\end{tabular}
\hfil
\caption{\footnotesize  Heavy right-handed  Majorana neutrino  masses  $M_{j}$
 for  degenerate models (IA,IB,IC), inverted models (IIA,IIB)
and normal hierarchical models (IIIA, IIIB),  with
$\tan^2\theta_{12}$=0.45, using neutrino mass matrices given in
Appendix A. The entry $(m,n)$ indicates the type of Dirac neutrino
mass matrix as down quark mass matrix (4,2), charged lepton mass
matrix (6,2) and up quark mass matrix (8,4) as explained in the text.  }
\end{table}
\begin{table}[tbp]
\begin{tabular}{llllllll} 
\hline
Type & (m,n) & $m_{\alpha\alpha}(eV)$&$\tilde{m_{1}}(eV)$&$\epsilon_{\alpha\alpha}$&$\epsilon$&$Y_{B1}$&$Y_{B3}$\\
\hline
IA &&1.13$\times 10^{-1}$&& 9.13$\times$$10^{-10}$& & &\\
IA &(4,2)&5.05$\times 10^{-1}$&1.12& 2.24$\times$$10^{-7}$&1.39$\times$$10^{-5}$&1.37$\times$$10^{-11}$&1.42$\times10^{-10}$\\
IA &&4.82$\times 10^{-1}$& &1.38$\times$$10^{-5}$& & &\\
\hline
IA &&1.32$\times 10^{-1}$&& 6.09$\times$$10^{-14}$& & &\\
IA &(6,2)&5.28$\times 10^{-1}$&1.19& 1.89$\times$$10^{-9}$&1.21$\times$$10^{-7}$&1.12$\times$$10^{-13}$&2.58$\times10^{-13}$\\
IA &&5.28$\times 10^{-1}$& &1.18$\times$$10^{-7}$& & &\\
\hline
IA &&1.32$\times 10^{-1}$&& 4.53$\times$$10^{-18}$& & &\\
IA &(8,4)&5.28$\times 10^{-1}$&1.19& 1.37$\times$$10^{-13}$&1.04$\times$$10^{-9}$&9.62$\times$$10^{-16}$&2.16$\times10^{-15}$\\
IA &&5.28$\times 10^{-1}$& &1.18$\times$$10^{-7}$& & &\\
\hline
\end{tabular}
\hfil
\begin{tabular}{llllllll} 
\hline
Type & (m,n) & $m_{\alpha\alpha}(eV)$&$\tilde{m_{1}}(eV)$&$\epsilon_{\alpha\alpha}$&$\epsilon$&$Y_{B1}$&$Y_{B3}$\\
\hline
IB &&3.97$\times 10^{-1}$&& 1.82$\times$$10^{-18}$& & &\\
IB &(4,2)&2.83$\times 10^{-9}$&0.3968& 9.02$\times$$10^{-19}$&2.76$\times$$10^{-14}$&7.66$\times$$10^{-20}$&1.09$\times10^{-11}$\\
IB &&2.78$\times 10^{-9}$& &2.76$\times$$10^{-14}$& & &\\
\hline
IB &&3.97$\times 10^{-1}$&& 1.19$\times$$10^{-22}$& & &\\
IB &(6,2)&2.79$\times 10^{-9}$&0.3968& 7.25$\times$$10^{-21}$&2.24$\times$$10^{-16}$&6.20$\times$$10^{-22}$&8.83$\times10^{-14}$\\
IB &&2.79$\times 10^{-9}$& &2.24$\times$$10^{-16}$& & &\\
\hline
IB &&3.96$\times 10^{-1}$&& 7.834$\times$$10^{-27}$& & &\\
IB &(8,4)&2.79$\times 10^{-9}$&0.3968& 4.72$\times$$10^{-25}$&1.81$\times$$10^{-18}$&5.02$\times$$10^{-24}$&7.15$\times10^{-16}$\\
IB &&2.79$\times 10^{-9}$& &1.81$\times$$10^{-7}$& & &\\
\hline
\end{tabular}
\hfil
\begin{tabular}{llllllll} 
\hline
Type & (m,n) & $m_{\alpha\alpha}(eV)$&$\tilde{m_{1}}(eV)$&$\epsilon_{\alpha\alpha}$&$\epsilon$&$Y_{B1}$&$Y_{B3}$\\
\hline
IC &&3.97$\times 10^{-1}$&& 1.31$\times$$10^{-16}$& & &\\
IC &(4,2)&2.78$\times 10^{-9}$&0.3968& 1.48$\times$$10^{-14}$&1.846$\times$$10^{-13}$&5.11$\times$$10^{-19}$&7.16$\times10^{-11}$\\
IC &&2.83$\times 10^{-9}$& &1.69$\times$$10^{-13}$& & &\\
\hline
IC &&3.97$\times 10^{-1}$&& 8.53$\times$$10^{-21}$& & &\\
IC &(6,2)&2.79$\times 10^{-9}$&0.3968& 1.19$\times$$10^{-17}$&1.47$\times$$10^{-16}$&4.08$\times$$10^{-22}$&5.81$\times10^{-14}$\\
IC &&2.79$\times 10^{-9}$& &1.35$\times$$10^{-16}$& & &\\
\hline
IC &&3.96$\times 10^{-1}$&& 4.61$\times$$10^{-23}$& & &\\
IC &(8,4)&2.79$\times 10^{-9}$&0.3968& 6.95$\times$$10^{-19}$&1.10$\times$$10^{-16}$&3.05$\times$$10^{-22}$&4.34$\times10^{-14}$\\
IC &&2.79$\times 10^{-9}$& &1.09$\times$$10^{-16}$& & &\\
\hline
\end{tabular}
\hfil
\caption{\footnotesize   Values of  CP asymmetry $\epsilon$ and $\epsilon_{\alpha\alpha}$ and  the baryon
 asymmetry $Y_{B1}$ and $Y_{B3}$ for degenerate models (IA, IB, IC)with for
  $\tan^{2}\theta_{12}$ =0.50 without and with flavour effects respectively, using light neutrino mass matrices given in Appendix A. The entry $(m,n)$ indicates the type of Dirac mass matrix as explained in the text. }
\end{table}
\begin{table}[tbp]
\begin{tabular}{llllllll} 
\hline
Type & (m,n) & $m_{\alpha\alpha}(eV)$&$\tilde{m_{1}}(eV)$&$\epsilon_{\alpha\alpha}$&$\epsilon$&$Y_{B1}$&$Y_{B3}$\\
\hline
IA &&3.57$\times 10^{-2}$&& 1.43$\times$$10^{-9}$& & &\\
IA &(4,2)&1.04$\times 10^{-1}$&2.38$\times10^{-1}$& 2.98$\times$$10^{-7}$&1.49$\times$$10^{-5}$&7.03$\times$$10^{-10}$&2.16$\times10^{-9}$\\
IA &&9.92$\times 10^{-2}$& &1.50$\times$$10^{-5}$& & &\\
\hline
IA &&3.57$\times 10^{-2}$&& 9.58$\times$$10^{-14}$& & &\\
IA &(6,2)&1.07$\times 10^{-1}$&2.50$\times10^{-1}$&2.50$\times$$10^{-9}$&1.31$\times$$10^{-7}$&5.76$\times$$10^{-12}$&1.34$\times10^{-11}$\\
IA &&1.07$\times 10^{-1}$& &1.28$\times$$10^{-7}$& & &\\
\hline
IA &&3.57$\times 10^{-2}$&& 7.52$\times$$10^{-18}$& & &\\
IA &(8,4)&1.07$\times 10^{-1}$&2.50$\times10^{-1}$& 1.91$\times$$10^{-13}$&1.16$\times$$10^{-9}$&5.12$\times$$10^{-14}$&1.19$\times10^{-13}$\\
IA &&1.07$\times 10^{-1}$& &1.16$\times$$10^{-9}$& & &\\
\hline
\end{tabular}
\hfil
\begin{tabular}{llllllll} 
\hline
Type & (m,n) & $m_{\alpha\alpha}(eV)$&$\tilde{m_{1}}(eV)$&$\epsilon_{\alpha\alpha}$&$\epsilon$&$Y_{B1}$&$Y_{B3}$\\
\hline
IB &&3.96$\times 10^{-1}$&& 1.68$\times$$10^{-18}$& & &\\
IB &(4,2)&2.65$\times 10^{-9}$&0.3964& 923$\times$$10^{-19}$&2.56$\times$$10^{-14}$&7.15$\times$$10^{-19}$&1.09$\times10^{-10}$\\
IB &&2.58$\times 10^{-9}$& &2.56$\times$$10^{-14}$& & &\\
\hline
IB &&3.96$\times 10^{-1}$&& 1.11$\times$$10^{-22}$& & &\\
IB &(6,2)&2.58$\times 10^{-9}$&0.3964& 7.50$\times$$10^{-21}$&2.06$\times$$10^{-16}$&5.76$\times$$10^{-21}$&8.84$\times10^{-13}$\\
IB &&2.58$\times 10^{-9}$& &2.08$\times$$10^{-16}$& & &\\
\hline
IB &&3.96$\times 10^{-1}$&& 7.27$\times$$10^{-27}$& & &\\
IB &(8,4)&2.58$\times 10^{-9}$&0.3964& 4.88$\times$$10^{-25}$&1.68$\times$$10^{-18}$&4.67$\times$$10^{-23}$&7.16$\times10^{-15}$\\
IB &&2.58$\times 10^{-9}$& &1.68$\times$$10^{-18}$& & &\\
\hline
\end{tabular}
\hfil
\begin{tabular}{llllllll} 
\hline
Type & (m,n) & $m_{\alpha\alpha}(eV)$&$\tilde{m_{1}}(eV)$&$\epsilon_{\alpha\alpha}$&$\epsilon$&$Y_{B1}$&$Y_{B3}$\\
\hline
IC &&3.96$\times 10^{-1}$&& 1.21$\times$$10^{-16}$& & &\\
IC &(4,2)&2.78$\times 10^{-9}$&0.3968& 1.37$\times$$10^{-14}$&1.85$\times$$10^{-13}$&5.12$\times$$10^{-18}$&7.16$\times10^{-10}$\\
IC &&2.83$\times 10^{-9}$& &1.69$\times$$10^{-13}$& & &\\
\hline
IC &&3.97$\times 10^{-1}$&& 8.53$\times$$10^{-21}$& & &\\
IC &(6,2)&2.79$\times 10^{-9}$&0.3968& 1.10$\times$$10^{-16}$&1.47$\times$$10^{-15}$&3.77$\times$$10^{-20}$&5.80$\times10^{-12}$\\
IC &&2.79$\times 10^{-9}$& &1.25$\times$$10^{-16}$& & &\\
\hline
IC &&3.96$\times 10^{-1}$&& 4.25$\times$$10^{-23}$& & &\\
IC &(8,4)&2.58$\times 10^{-9}$&0.3968& 6.41$\times$$10^{-19}$&1.02$\times$$10^{-16}$&2.82$\times$$10^{-21}$&4.34$\times10^{-13}$\\
IC &&2.58$\times 10^{-9}$& &1.01$\times$$10^{-16}$& & &\\
\hline
\end{tabular}
\hfil
\caption{\footnotesize   Values of  CP asymmetry $\epsilon$ and $\epsilon_{\alpha\alpha}$ and  the baryon
 asymmetry $Y_{B1}$ and $Y_{B3}$ for degenerate models (IA, IB, IC)with for
  $\tan^{2}\theta_{12}$ =0.45 without and with flavour effects respectively, using light neutrino mass matrices given in Appendix A. The entry $(m,n)$ indicates the type of Dirac mass matrix as explained in the text.  }
\end{table}

\begin{table}[tbp]
\begin{tabular}{llllllll} 
\hline
Type & (m,n) & $m_{\alpha\alpha}(eV)$&$\tilde{m_{1}}(eV)$&$\epsilon_{\alpha\alpha}$&$\epsilon$&$Y_{B1}$&$Y_{B3}$\\
\hline
IIA &&4.95$\times 10^{-2}$&& 5.86$\times$$10^{-19}$& & &\\
IIA &(4,2)&1.22$\times 10^{-6}$&4.95$\times10^{-1}$& 7.46$\times$$10^{-15}$&9.37$\times$$10^{-13}$&2.08$\times$$10^{-17}$&8.14$\times10^{-13}$\\
IIA &&1.22$\times 10^{-6}$& &9.30$\times$$10^{-13}$& & &\\
\hline
IIA &&4.95$\times 10^{-2}$&& 3.83$\times$$10^{-23}$& & &\\
IIA &(6,2)&1.22$\times 10^{-6}$&4.95$\times10^{-2}$ & 5.99$\times$$10^{-17}$&7.53$\times$$10^{-15}$&1.67$\times$$10^{-19}$&6.83$\times10^{-15}$\\
IIA &&1.21$\times 10^{-6}$& &7.47$\times$$10^{-15}$& & &\\
\hline
IIA &&4.95$\times 10^{-2}$&& 2.61$\times$$10^{-27}$& & &\\
IIA &(8,4)&1.21$\times 10^{-6}$&4.95$\times10^{-2}$ & 4.02$\times$$10^{-21}$&6.19$\times$$10^{-17}$&1.38$\times$$10^{-21}$&5.63$\times10^{-17}$\\
IIA &&1.21$\times 10^{-6}$& &6.19$\times$$10^{-17}$& & &\\
\hline
\end{tabular}
\hfil
\begin{tabular}{llllllll} 
\hline
Type & (m,n) & $m_{\alpha\alpha}(eV)$&$\tilde{m_{1}}(eV)$&$\epsilon_{\alpha\alpha}$&$\epsilon$&$Y_{B1}$&$Y_{B3}$\\
\hline
IIB &&1.61$\times 10^{-2}$&& 1.36$\times$$10^{-10}$& & &\\
IIB &(4,2)&6.22$\times 10^{-2}$& 1.42$\times10^{-1}$& 2.94$\times$$10^{-8}$&5.73$\times$$10^{-6}$&4.42$\times$$10^{-11}$&9.82$\times10^{-11}$\\
IIB &&6.42$\times 10^{-2}$& &5.70$\times$$10^{-6}$& & &\\
\hline
IIB &&1.61$\times 10^{-2}$&& 7.98$\times$$10^{-15}$& & &\\
IIB &(6,2)&6.78$\times 10^{-2}$&1.52$\times10^{-1}$ & 7.99$\times$$10^{-15}$&4.28$\times$$10^{-8}$&3.10$\times$$10^{-13}$&6.94$\times10^{-13}$\\
IIB &&6.78$\times 10^{-2}$& &4.26$\times$$10^{-8}$& & &\\
\hline
IIB && 1.61$\times 10^{-2}$&& 5.94$\times$$10^{-19}$& & &\\
IIB &(8,4)&6.78$\times 10^{-2}$&1.52$\times10^{-1}$ & 1.64$\times$$10^{-14}$&3.88$\times$$10^{-10}$&2.81$\times$$10^{-13}$&6.29$\times10^{-13}$\\
IIB &&6.78$\times 10^{-2}$& &3.88$\times$$10^{-10}$& & &\\
\hline
\end{tabular}
\hfil
\caption{\footnotesize  Values of  CP asymmetry $\epsilon$ and $\epsilon_{\alpha\alpha}$ and  the baryon
 asymmetry $Y_{B1}$ and $Y_{B3}$ for inverted hierarchical models (IIA, IIB)
  without and with flavour effects respectively for $\tan^{2}\theta_{12}$ =0.50, using light neutrino mass matrices
 given in Appendix A. The entry $(m,n)$ indicates the type of Dirac
 mass matrix as explained in the text. }
\end{table}
\begin{table}[tbp]
\begin{tabular}{llllllll} 
\hline
Type & (m,n) & $m_{\alpha\alpha}(eV)$&$\tilde{m_{1}}(eV)$&$\epsilon_{\alpha\alpha}$&$\epsilon$&$Y_{B1}$&$Y_{B3}$\\
\hline
IIA &&4.94$\times 10^{-2}$&& 6.76$\times$$10^{-19}$& & &\\
IIA &(4,2)&1.56$\times 10^{-6}$&4.95$\times10^{-2}$& 9.07$\times$$10^{-15}$&1.12$\times$$10^{-12}$&2.49$\times$$10^{-16}$&7.90$\times10^{-12}$\\
IIA &&1.56$\times 10^{-6}$& &1.13$\times$$10^{-12}$& & &\\
\hline
IIA &&4.94$\times 10^{-2}$&& 4.41$\times$$10^{-23}$& & &\\
IIA &(6,2)&1.55$\times 10^{-6}$&4.95$\times10^{-2}$ & 7.28$\times$$10^{-17}$&9.00$\times$$10^{-15}$&2.00$\times$$10^{-18}$&6.34$\times10^{-14}$\\
IIA &&1.55$\times 10^{-6}$& &9.08$\times$$10^{-15}$& & &\\
\hline
IIA &&4.94$\times 10^{-2}$&& 3.04$\times$$10^{-27}$& & &\\
IIA &(8,4)&1.55$\times 10^{-6}$&4.95$\times10^{-2}$ & 4.89$\times$$10^{-21}$&7.53$\times$$10^{-17}$&1.67$\times$$10^{-20}$&5.35$\times10^{-16}$\\
IIA &&1.54$\times 10^{-6}$& &7.53$\times$$10^{-17}$& & &\\
\hline
\end{tabular}
\hfil
\begin{tabular}{llllllll} 
\hline
Type & (m,n) & $m_{\alpha\alpha}(eV)$&$\tilde{m_{1}}(eV)$&$\epsilon_{\alpha\alpha}$&$\epsilon$&$Y_{B1}$&$Y_{B3}$\\
\hline
IIB &&1.99$\times 10^{-2}$&& 9.04$\times$$10^{-11}$& & &\\
IIB &(4,2)&5.74$\times 10^{-2}$& 1.36$\times10^{-1}$& 2.13$\times$$10^{-8}$&4.02$\times$$10^{-6}$&3.25$\times$$10^{-10}$&7.53$\times10^{-10}$\\
IIB &&5.87$\times 10^{-2}$& &4.00$\times$$10^{-6}$& & &\\
\hline
IIB &&1.98$\times 10^{-2}$&& 5.92$\times$$10^{-15}$& & &\\
IIB &(6,2)&6.14$\times 10^{-2}$&1.43$\times10^{-1}$ & 1.78$\times$$10^{-10}$&3.33$\times$$10^{-8}$&2.57$\times$$10^{-12}$&5.96$\times10^{-12}$\\
IIB &&6.14$\times 10^{-2}$& &3.31$\times$$10^{-8}$& & &\\
\hline
IIB &&1.98$\times 10^{-2}$&& 3.97$\times$$10^{-19}$& & &\\
IIB &(8,4)&6.14$\times 10^{-2}$&1.42$\times10^{-1}$ & 1.78$\times$$10^{-14}$&2.71$\times$$10^{-10}$&2.09$\times$$10^{-14}$&4.86$\times10^{-14}$\\
IIB &&6.14$\times 10^{-2}$& &2.71$\times$$10^{-10}$& & &\\
\hline
\end{tabular}
\hfil
\caption{\footnotesize  Values of  CP asymmetry $\epsilon$ and $\epsilon_{\alpha\alpha}$ and  the baryon
 asymmetry $Y_{B1}$ and $Y_{B3}$ for inverted hierarchical models (IIA, IIB)
  without and with flavour effects respectively for $\tan^{2}\theta_{12}$ =0.45, using light neutrino mass matrices
 given in Appendix A. The entry $(m,n)$ indicates the type of Dirac
 mass matrix as explained in the text.   }
\end{table}
\begin{table}[tbp]
\begin{tabular}{llllllll} 
\hline
Type & (m,n) & $m_{\alpha\alpha}(eV)$&$\tilde{m_{1}}(eV)$&$\epsilon_{\alpha\alpha}$&$\epsilon$&$Y_{B1}$&$Y_{B3}$\\
\hline
IIIA &&2.17$\times 10^{-4}$&& 7.44$\times$$10^{-9}$& & &\\
IIIA &(4,2)&4.13$\times 10^{-2}$&4.38$\times10^{-2}$& 1.19$\times$$10^{-6}$&3.43$\times$$10^{-5}$&8.65$\times$$10^{-10}$&1.79$\times10^{-8}$\\
IIIA &&2.05$\times 10^{-3}$& &3.31$\times$$10^{-5}$& & &\\
\hline
IIIA &&2.00$\times 10^{-5}$&& 1.79$\times$$10^{-12}$& & &\\
IIIA &(6,2)&3.15$\times 10^{-1}$&5.80$\times10^{-1}$ & 3.24$\times$$10^{-7}$&3.70$\times$$10^{-5}$&7.01$\times$$10^{-11}$&1.53$\times10^{-10}$\\
IIIA &&2.64$\times 10^{-1}$& &3.66$\times$$10^{-5}$& & &\\
\hline
IIIA &&1.99$\times 10^{-5}$&& 1.23$\times$$10^{-16}$& & &\\
IIIA &(8,4)&3.16$\times 10^{-1}$&5.81$\times10^{-1}$ & 2.19$\times$$10^{-11}$&3.05$\times$$10^{-7}$&5.77$\times$$10^{-13}$&1.27$\times10^{-12}$\\
IIIA &&2.64$\times 10^{-1}$& &3.05$\times$$10^{-7}$& & &\\
\hline
\end{tabular}
\hfil
\begin{tabular}{llllllll} 
\hline
Type & (m,n) & $m_{\alpha\alpha}(eV)$&$\tilde{m_{1}}(eV)$&$\epsilon_{\alpha\alpha}$&$\epsilon$&$Y_{B1}$&$Y_{B3}$\\
\hline
IIIB &&2.29$\times 10^{-4}$&& 7.90$\times$$10^{-9}$& & &\\
IIIB &(4,2)&4.18$\times 10^{-2}$& 4.52$\times10^{-2}$& 1.35$\times$$10^{-6}$&4.82$\times$$10^{-5}$&1.17$\times$$10^{-9}$&1.62$\times10^{-8}$\\
IIIB &&3.20$\times 10^{-3}$& &4.69$\times$$10^{-5}$& & &\\
\hline
IIIB &&2.11$\times 10^{-5}$&& 1.50$\times$$10^{-12}$& & &\\
IIIB &(6,2)&3.30$\times 10^{-1}$&6.11$\times10^{-1}$ & 2.96$\times$$10^{-7}$&3.40$\times$$10^{-5}$&6.13$\times$$10^{-11}$&1.33$\times10^{-10}$\\
IIIB &&2.81$\times 10^{-1}$& &3.37$\times$$10^{-5}$& & &\\
\hline
IIIB &&2.10$\times 10^{-5}$&& 1.02$\times$$10^{-16}$& & &\\
IIIB &(8,4)&3.31$\times 10^{-1}$&6.13$\times10^{-1}$ & 2.00$\times$$10^{-11}$&2.81$\times$$10^{-7}$&5.04$\times$$10^{-13}$&1.10$\times10^{-12}$\\
IIIB &&2.82$\times 10^{-1}$& &2.80$\times$$10^{-7}$& & &\\
\hline
\end{tabular}
\hfil
\caption{\footnotesize   Values of  CP asymmetry $\epsilon$ and $\epsilon_{\alpha\alpha}$ and  the baryon
 asymmetry $Y_{B1}$ and $Y_{B3}$ for normal hierarchical models (IIIA, IIIB)
  without and with flavour effects respectively for $\tan^{2}\theta_{12}$ =0.50, using mass matrices
 given in Appendix A. The entry $(m,n)$ indicates the type of Dirac
 mass matrix as explained in the text. }
\end{table}
\begin{table}[tbp]
\begin{tabular}{llllllll} 
\hline
Type & (m,n) & $m_{\alpha\alpha}(eV)$&$\tilde{m_{1}}(eV)$&$\epsilon_{\alpha\alpha}$&$\epsilon$&$Y_{B1}$&$Y_{B3}$\\
\hline
IIIA &&8.94$\times 10^{-4}$&& 8.64$\times$$10^{-8}$& & &\\
IIIA &(4,2)&3.01$\times 10^{-2}$&4.28$\times10^{-2}$& 1.24$\times$$10^{-5}$&2.22$\times$$10^{-4}$&5.71$\times$$10^{-8}$&2.00$\times10^{-7}$\\
IIIA &&1.19$\times 10^{-2}$& &2.10$\times$$10^{-4}$& & &\\
\hline
IIIA &&8.78$\times 10^{-4}$&& 8.34$\times$$10^{-12}$& & &\\
IIIA &(6,2)&3.44$\times 10^{-2}$&6.95$\times10^{-2}$ & 1.61$\times$$10^{-7}$&4.34$\times$$10^{-6}$&6.86$\times$$10^{-10}$&1.39$\times10^{-9}$\\
IIIA &&3.42$\times 10^{-2}$& &4.18$\times$$10^{-6}$& & &\\
\hline
IIIA &&8.78$\times 10^{-4}$&& 1.04$\times$$10^{-15}$& & &\\
IIIA &(8,4)&3.44$\times 10^{-2}$&6.96$\times10^{-2}$ & 1.91$\times$$10^{-11}$&5.06$\times$$10^{-8}$&7.99$\times$$10^{-12}$&1.62$\times10^{-11}$\\
IIIA &&3.44$\times 10^{-2}$& &5.05$\times$$10^{-8}$& & &\\
\hline
\end{tabular}
\hfil
\begin{tabular}{llllllll} 
\hline
Type & (m,n) & $m_{\alpha\alpha}(eV)$&$\tilde{m_{1}}(eV)$&$\epsilon_{\alpha\alpha}$&$\epsilon$&$Y_{B1}$&$Y_{B3}$\\
\hline
IIIB &&2.09$\times 10^{-4}$&& 7.18$\times$$10^{-9}$& & &\\
IIIB &(4,2)&3.99$\times 10^{-2}$& 4.19$\times10^{-2}$& 1.13$\times$$10^{-6}$&3.09$\times$$10^{-5}$&8.13$\times$$10^{-9}$&1.85$\times10^{-7}$\\
IIIB &&1.77$\times 10^{-3}$& &2.98$\times$$10^{-5}$& & &\\
\hline
IIIB &&1.93$\times 10^{-5}$&& 1.85$\times$$10^{-12}$& & &\\
IIIB &(6,2)&3.04$\times 10^{-1}$&5.59$\times10^{-1}$ & 3.29$\times$$10^{-7}$&3.74$\times$$10^{-5}$&7.37$\times$$10^{-10}$&1.62$\times10^{-9}$\\
IIIB &&2.54$\times 10^{-1}$& &3.71$\times$$10^{-5}$& & &\\
\hline
IIIB &&1.93$\times 10^{-5}$&& 1.26$\times$$10^{-16}$& & &\\
IIIB &(8,4)&3.06$\times 10^{-1}$&5.60$\times10^{-1}$ & 2.22$\times$$10^{-11}$&3.09$\times$$10^{-7}$&6.06$\times$$10^{-12}$&1.13$\times10^{-11}$\\
IIIB &&2.55$\times 10^{-1}$& &3.09$\times$$10^{-7}$& & &\\
\hline
\end{tabular}
\hfil
\caption{\footnotesize  Values of  CP asymmetry $\epsilon$ and $\epsilon_{\alpha\alpha}$ and  the baryon
 asymmetry $Y_{B1}$ and $Y_{B3}$ for normal hierarchical models (IIIA, IIIB)
  without and with flavour effects respectively for $\tan^{2}\theta_{12}$ =0.45, using mass matrices
 given in Appendix A. The entry $(m,n)$ indicates the type of Dirac
 mass matrix as explained in the text.  }
\end{table}
 \section{Results and Discussion}
The numerical predictions on $\bigtriangleup m^2_{21}$ and
$\bigtriangleup m^2_{23}$ of these seven neutrino mass models $m_{LL}$ under
consideration in Appendix A, are presented in Table 1 and Table 2 for $tan^2\theta_{12}=$ 0.5 and 0.45 respectively. They obey $\mu-\tau$ symmetry and predict
tribimaximal mixings as expected. In Table 3 and Table 4 the three heavy right-handed
neutrino masses are extracted from the right-handed Majorana mass
matrices so constructed through the inverse seesaw formula, for three
choices of diagonal Dirac neutrino mass matrices. The corresponding 
baryon 
asymmetry $Y_B$ are estimated following sections 2 and 3, for degenerate
model(IA,IB,IC), inverted hierarchical models(IIA,IIB) and normal
hierarchical
 models(IIIA, IIIB) respectively as indicated in the Tables (5-10).

In these calculations we have focussed on two issues :(i) dependence
of $Y_{B}$
 on lepton flavours, (ii) dependence of $Y_{B}$ on type of Dirac
 neutrino mass matrix.
 We have found strong dependence on the type of Dirac neutrino mass
 matrix,
 where down-quark type mass matrix
 $(\lambda^{4},\lambda^{2},1)v$ 
leading to highest contribution and charged lepton mass matrix 
$(\lambda^{6},\lambda^{2},1)v$ and up-quark type mass
 matrix $(\lambda^{8},\lambda^{4},1)v$  in decreasing order
 with factor of 100, in all cases. The enhancement in flavoured
 leptogenesis is also a common feature for all cases, and such
 enhancement
 is also dependent on the type of the Dirac neutrino mass matrix.

 Both normal hierarchical models(IIIA,IIIB) predict good results
 consistent with observations for the case with Dirac neutrino mass
 matrix in Table (9-10). Inverted hierarchical model(IIB) with
 $(m,n)=(4,2)$
 also leads to acceptable results and it is not yet ruled out. 
However inverted hierarchical model(IIA) is completely ruled out as
seen in Table (7-8).
 The degenerate models (IA,IB,IC) with $(m,n)=(4,2)$ in the case of 
flavoured leptogenesis still show reasonable prediction in Table (5-6).

 The present analysis is  extended for $\mu-\tau$ symmetric mass
 matrices $m_{LL}$ with $tan^{2}\theta_{12}=0.45$ . The analysis indicates an enhancement in the baryon asymmetry by a factor of one. In some left-right 
symmetric SO(10) GUT, Dirac neutrino mass matrix is considered as
charged lepton
 type mass matrix. In such condition only normal hierarchical
 model leads to good prediction consistent with data.

The present analysis if considered as an additional criteria for the 
discrimination of  neutrino mass models, may lead to normal
hierarchical model
 as the most favourable choice of nature.
 This conclusion is consistent with other conditions such as 
 stability criteria under quantum radiative corrections in MSSM. Moreover,the normal hierarchical model also leads to a good prediction with the Type II seesaw formula as well \cite{ref AZN15}. 
\newpage
\section*{Appendix A: Possible patterns of neutrino mass models obeying $\mu-\tau$ symmetry with two parameters $\epsilon$ and $\eta$}
 Left-handed Majorana neutrino mass matrices which obey
   $\mu-\tau $ symmetry\cite{ref NZM20,ref deg} have the following form
\[ m_{LL} = \left( \begin{array}{ccc}
 X  & Y  & Y \\
 Y &  Z & W \\
 Y &  W & Z \end{array} \right)m_{o} \]\\
This predicts an arbitrary solar mixing angle $\tan
2\theta_{12}=|\frac{2\sqrt{2}Y}{(X-Z-W)}|$, while the predictions on
atmospheric mixing angle is maximal $(\theta_{23}=\pi/4)$ and Chooz angle is 
zero. We parametrise the mass matrices (with only two parameters $\epsilon$ and $\eta$ )
whereby the solar mixing is fixed at  tribimaximal mixings for all
possible patterns of neutrino mass models:\\
1.\underline{Deg Type A (IA)}($m_i=m_1,-m_2,m_3$)\\
 \[ m_{LL} = \left( \begin{array}{ccc}
 \epsilon-2\eta & -\epsilon  & -\epsilon \\
-\epsilon &  \frac{1}{2}-\eta & -\frac{1}{2}-\eta \\
  -\epsilon &  -\frac{1}{2}-\eta & \frac{1}{2}-\eta \end{array} \right)m_{o} \]\\
with input values: $\epsilon$=0.66115, $\eta$=0.16535, $m_{o}=0.4 eV$.\\
2.\underline{Deg Type B (IB)}($m_i=m_1,m_2,m_3$)\\
 \[ m_{LL} = \left( \begin{array}{ccc}
 1-\epsilon-2\eta & \epsilon  & \epsilon \\
\epsilon &  1-\eta & -\eta \\
  \epsilon & -\eta & 1-\eta \end{array} \right)m_{o} \]\\ 
with input values: $\epsilon$=8.314$\times$$10^{-5}$,$\eta$=0.00395,$m_{o}$=0.4eV.\\
3.\underline{Deg Type C (IC)}($m_i=m_1,m_2,-m_3$)\\
 \[ m_{LL} = \left( \begin{array}{ccc}
 1-\epsilon-2\eta & \epsilon  & \epsilon \\
\epsilon &  -\eta & 1-\eta \\
  \epsilon & 1-\eta & -\eta \end{array} \right)m_{o} \]\\ 
with input values: $\epsilon$=8.314$\times$$10^{-5}$,$\eta$=0.00395,$m_{o}$=0.4eV.\\
4:\underline{Inverted Hierarchical mass matrix}
{\bf  with $m_3\neq 0$:}
\[ m_{LL}(IH)=
 \left(\begin{array}{ccc}
1 -2\epsilon  &  -\epsilon  &   -\epsilon \\
  -\epsilon   &  1/2  &   1/2-\eta  \\
 -\epsilon  &  1/2-\eta  &  1/2
 \end{array}\right)m_0.\] \\
Inverted Hierarchy with even CP parity in the first two mass
 eigenvalues (IIA) $(m_1=m_1,m_2,m_3)$: $\eta /\epsilon$=1.0,$\eta$=0.0048,$m_{0}=0.05 eV$.\\
 Inverted Hierarchy with odd CP parity in the first two mass
 eigenvalues (IIB) $(m_i=m_1,-m_2,m_3)$: $\eta /\epsilon$=1.0,$\eta$=0.6607,$m_{0}=0.05 eV$.\\
5:.\underline{Normal Hierarchical mass matrix}
{\bf  Case (i)  with $m(1,1)=X \neq 0$} type- IIIA:
\[m_{LL}(NH)=
 \left(\begin{array}{ccc}
 -\eta  &  -\epsilon  &   -\epsilon \\
  -\epsilon   &  1-\epsilon  &   -1  \\
 -\epsilon  &  -1  &  1-\epsilon
 \end{array}\right)m_0 \] \\
with input values: $\eta /\epsilon$=0.0,$\epsilon$=0.175,$m_{0}=0.029 eV$.\\
6:\underline{Normal Hierarchical mass matrix}
{\bf Case (ii)  with $m(1,1)=X=0$; type- IIIB}:
\[ m_{LL}(NH)=
 \left(\begin{array}{ccc}
 0  &  -\epsilon  &   -\epsilon \\
  -\epsilon   &  1-\epsilon  &   -1+\eta  \\
 -\epsilon  &  -1+\eta  &  1-\epsilon
 \end{array}\right)m_0 \] \\
with input values: $\eta /\epsilon$=0.0,
$\epsilon$=0.164,$m_{0}=0.028eV$.
The textures of mass matrices for degenerate(IA, IB, IC), inverted hierarchy (IIA, IIB) as well as normal hierarchy (IIIA, IIIB) have the potential to decrease the solar mixing angle from the tribimaximal value, without sacrificing
$\mu-\tau$ symmetry.  This is possible through the identification of
'flavour twister' $\eta/\epsilon \neq 0$ \cite{ref NZM20,ref deg}. The values of $\epsilon$ and $\eta$ for $tan^{2}\theta_{12}$=0.45 are collected from \cite{ref NZM20,ref deg}.  
\section*{Acknowledgement}
One of us HZD would like to thank CSIR for the Senior Research
Fellowship as financial assistance for carrying out her research work.

\end{document}